\documentclass{article}
\usepackage{spconf,amsmath,graphicx,hyperref}
\usepackage{float}
\usepackage{algorithm}
\usepackage{algpseudocode}
\usepackage{amsfonts}
\usepackage{booktabs, multirow} 
\usepackage{enumitem}
\setlist{nosep}

\title{MedCondDiff: Lightweight, Robust, Semantically Guided Diffusion \\ for Medical Image Segmentation}

\twoauthors
 {Ruirui Huang$^{*,\dagger}$}
	{$^*$Johns Hopkins University\\
    Department of Computer Science\\
	Baltimore, MD 21218, USA}
 {Jiacheng Li$^{\dagger}$}
	{$^\dagger$ Johns Hopkins University\\
	Laboratory of Computational Sensing and Robotics\\
	Baltimore, MD 21218, USA}
\begin{document}
%
\maketitle
© 2025 IEEE. Personal use of this material is permitted. Permission from IEEE must be obtained for all other uses, in any current or future media, including reprinting/republishing this material for advertising or promotional purposes, creating new collective works, for resale or redistribution to servers or lists, or reuse of any copyrighted component of this work in other works.
\begin{abstract}
We introduce MedCondDiff, a diffusion-based framework for multi-organ medical image segmentation that is efficient and anatomically grounded. The model conditions the denoising process on semantic priors extracted by a Pyramid Vision Transformer (PVT) backbone, yielding a semantically guided and lightweight diffusion architecture. This design improves robustness while reducing both inference time and VRAM usage compared to conventional diffusion models. Experiments on multi-organ, multi-modality datasets demonstrate that MedCondDiff delivers competitive performance across anatomical regions and imaging modalities, underscoring the potential of semantically guided diffusion models as an effective class of architectures for medical imaging tasks. Code is available at \url{https://github.com/ruiruihuangannie/MedCondDiff}.
\end{abstract}
\begin{keywords}
Computer Vision, Medical Imaging, Diffusion Model, Image Segmentation, Deep Learning
\end{keywords}
\section{Introduction}
\label{sec:intro}
Accurate medical image segmentation underpins modern clinical workflows, enabling diagnosis, treatment planning, and surgical navigation through reproducible analysis of scans \cite{medseg-survey-1, medseg-survey-2, medseg-survey-3, medseg-survey-4}. Deep learning frameworks for segmentation are primarily based on Convolutional Neural Networks (CNN) \cite{2015unet, 2016vnet}, Vision Transformers (ViT) \cite{vit2021, bao2023worthwordsvitbackbone}, or their hybrids \cite{2024vit-cnn-survey}. CNNs are limited by local receptive fields and struggle with long-range dependencies, while ViTs capture global context but demand large datasets and high computational cost.

\begin{figure}[!ht]
    \centering
    \includegraphics[width=\linewidth]{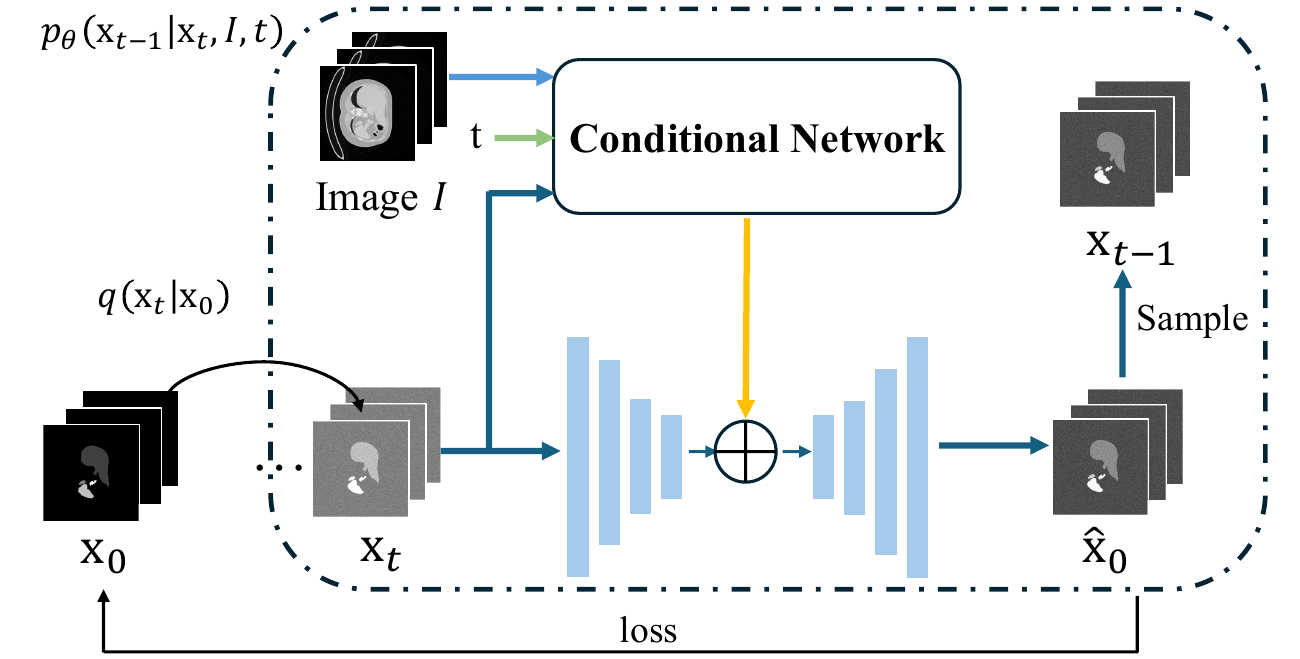}
    \caption{MedCondDiff Training framework. The Conditional Network (‘Adapter’) extracts features and injects them into the denoising network to enhance mask prediction.}
    \label{fig-1}
\end{figure}

Recently, Diffusion probabilistic models (DPMs) offer a promising alternative by framing segmentation as a reverse-time stochastic process \cite{dpm, ho2020denoising, 2022ddim}. They generate high-quality outputs but, in their vanilla form, lack constraints on organ presence or structure and perform poorly.

We present MedCondDiff, a conditional diffusion-based segmentation framework that incorporates structured semantic priors through a lightweight adapter mechanism. Building on adapter-based conditioning modules in vision–language tasks \cite{2023t2i, 2023ip}, we extend this idea to the medical domain by designing a modular conditional adapter for diffusion probabilistic models (DPMs).

MedCondDiff generates segmentation masks by injecting domain-specific information into the diffusion process, guiding predictions toward anatomical fidelity. A UNet-based denoising network iteratively refines multi-organ masks under conditional guidance, while a Pyramid Vision Transformer (PVT) backbone provides rich anatomical context. Finally, we employ a iterative prediction fusion strategy that fuses intermediate predictions to enhance mask consistency.

\begin{figure*}[!t]
    \centering
    \includegraphics[width=\textwidth]{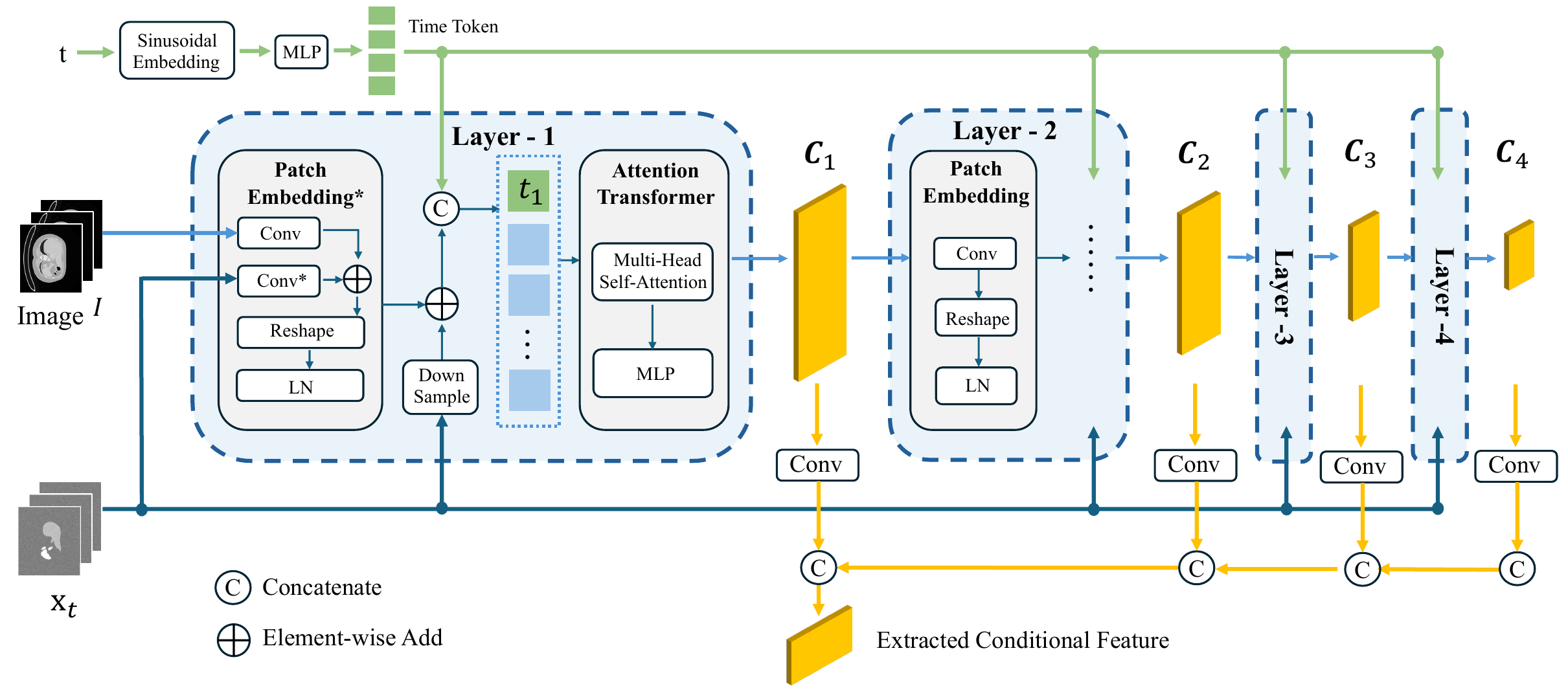}
    \caption{MedCondDiff Conditioning framework. This ‘Adapter’ extracts features and injects them into the denoising network to enhance mask prediction. In the first layer of conditional network, block labeled with $'*'$ processes the noised image $\mathbf{x_t}$ and combines it with regular embedding.}
    \label{fig-2}
\end{figure*}

Our main contributions are:
\begin{enumerate}
    \item A generalized adapter framework with formal formulation, enabling modular conditioning of diffusion models with structured priors.
    \item A lightweight PVT-guided conditional diffusion architecture for medical image segmentation, integrating long-range dependencies and hierarchical features.
    \item Empirical validation on two public datasets, showing competitive accuracy, reduced memory, and runtime.
\end{enumerate}

\section{Methodology}
\label{sec:method}

\subsection{Denoising Diffusion Probabilistic Models (DDPMs)}
DDPMs are generative models that define a Markov diffusion process \( q(\mathbf{x}_{1:T} | \mathbf{x}_0) \), which gradually corrupts data \( \mathbf{x}_0 \) into noise, and a reverse process \( p_\theta(\mathbf{x}_{0:T}) \) that learns to reconstruct data from noise. The forward pass is \cite{ho2020denoising}:
\begin{equation}
q(\mathbf{x}_{1:T}|\mathbf{x}_0) = \prod_{t=1}^T q(\mathbf{x}_t | \mathbf{x}_{t-1})    
\end{equation}
\begin{equation}
\quad q(\mathbf{x}_t|\mathbf{x}_{t-1}) := \mathcal{N}(\mathbf{x}_t\sqrt{1 - \beta_t}\mathbf{x}_{t-1}, \beta_t \mathbf{I})
\end{equation}
Where \(\beta_t \in (0, 1), t \in \{1, 2, \ldots, T\}\) is the noise schedule that regulates variance in each step. Given this forward process, the reverse process attempts to invert the noise via:
\begin{equation}
p_\theta(\mathbf{x}_{0:T}) = p(\mathbf{x}_T)\prod_{t=1}^T p_\theta(\mathbf{x}_{t-1} | \mathbf{x}_t)
\end{equation}
\begin{equation}
\quad p_\theta(\mathbf{x}_{t-1} | \mathbf{x}_t) := \mathcal{N}(\mathbf{x}_{t-1}\boldsymbol{\mu}_\theta(\mathbf{x}_t, t), \boldsymbol{\Sigma}_\theta(\mathbf{x}_t, t))
\end{equation}

The training objective maximizes the evidence lower bound (ELBO), which can be decomposed into KL divergences between the approximate and true posteriors:
\begin{equation}
\mathcal{L} = \mathbb{E}_{q(\mathbf{x}_{0:T})} \left[ \sum_{t=1}^T D_{KL}(q(\mathbf{x}_{t-1} | \mathbf{x}_t, \mathbf{x}_0) || p_\theta(\mathbf{x}_{t-1} | \mathbf{x}_t)) \right]
\end{equation}

\subsection{Pyramid Vision Transformer (PVT)}
PVT \cite{wang2021pvt} introduces a hierarchical transformer architecture tailored for dense prediction tasks. Unlike vanilla ViTs with fixed token resolution, PVT produces multi-scale feature maps:
\begin{equation}
\{\mathbf{C}_1, \mathbf{C}_2, \ldots, \mathbf{C}_S\}, \quad \mathbf{C}_s \in \mathbb{R}^{\text{H}_s \times \text{W}_s \times \text{Ch}_s},
\end{equation}
where spatial dimensions $\text{H}_s, \text{W}_s$ decrease stage-wise, forming a pyramid similar to CNN backbones. Overlapping patch embeddings improve local continuity and richer features.

To reduce the quadratic complexity of self-attention, PVT applies a spatial reduction function $\mathcal{R}(\cdot)$ to keys and values, lowering attention complexity from $\mathcal{O}(N^2)$ to approximately $\mathcal{O}(N \cdot \frac{N}{r^2})$, where $r$ is the spatial reduction ratio.

Formally, the spatial-reduction attention at stage $s$ can be abstracted as:
\begin{equation}
\mathrm{Attention}_s = \mathrm{Softmax}\left(\frac{\mathbf{Q}_s \cdot \mathcal{R}(\mathbf{K}_s)^\top}{\sqrt{d}}\right) \cdot \mathcal{R}(\mathbf{V}_s).
\end{equation}
where $d$ represents the dimension of each attention head. 

Through multi-stage transformer blocks , PVT produces hierarchical embeddings that effectively balance global context and local details, making it well-suited as a backbone for segmentation and other dense vision tasks.

\subsection{Adapter-Based Conditioning for Segmentation}
To formalize our approach, we develop a unified network– adapter framework that organizes conditional modeling strategies across tasks such as image generation, visual correspondence, and segmentation. The main network is shown in Fig. \ref{fig-1}, and the adapter module (Fig. \ref{fig-2}) can be integrated via concatenation, attention, or element-wise addition. This abstraction highlights the adapter’s modularity for task-specific conditioning without altering the core architecture.

Within this paradigm, prior works demonstrate the flexibility of adapters: T2I-Adapter \cite{2023t2i} injects structural priors without retraining the base network, ControlNet \cite{2023ControlNet} guides denoising through an auxiliary branch, and ConSept \cite{dong2024consept} introduces lightweight attention-based adapters for continual segmentation. Following this principle, we design a PVT-based adapter that supplies hierarchical semantic context to the diffusion process through additive conditioning, reducing anatomically implausible outputs and promoting structurally consistent predictions.

\subsection{Training and Sampling Strategy}
Hierarchical, multi-scale features extracted by the PVT adapter serve as semantic priors, \( \mathbf{c} = \mathcal{E}_{\text{PVT}}(I, \mathbf{x}_t, t)\) that conditions the denoising network at each step. The reverse process predicts the clean mask $\hat{\mathbf{x}}_0$ using high-level embeddings from a pre-trained PVT backbone, as in Alg \ref{Alg1}, \ref{Alg2}.
 
During training, the forward diffusion process gradually corrupts the clean image \(\mathbf{x}_0\), by adding Gaussian noise following a variance schedule \(\beta_t\), producing a noisy sample \(\mathbf{x}_t\) at timestep \(t\) \cite{2022ddim}:
\begin{equation}
\mathbf{x}_t = \sqrt{\bar{\alpha}_t} \mathbf{x}_0 + \sqrt{1 - \bar{\alpha}_t} \epsilon, \quad \epsilon \sim \mathcal{N}(0, \mathbf{I}).
\end{equation}
where $\bar{\alpha}_t = \prod_{k=1}^t(1-\beta_k)$. 
The reverse denoising process aims to iteratively reconstruct \(\mathbf{x}_0\) from \(\mathbf{x}_t\), conditioned on \(c\). The reverse mean function is defined as \cite{ho2020denoising}:
\begin{equation}
\boldsymbol{\mu}_\theta(\mathbf{x}_t, t,\mathbf{c} ) = \frac{\sqrt{\alpha_t} (1 - \bar{\alpha}_{t-1})}{1 - \bar{\alpha}_t} \mathbf{x}_t + \frac{\sqrt{\bar{\alpha}_{t-1}} (1 - \alpha_t)}{1 - \bar{\alpha}_t} \mathbf{\hat{x}}_0(\mathbf{x}_t, \mathbf{c}, t)
\end{equation}
where the predicted image $\mathbf{\hat{x}}_0$ is given by:
\begin{equation}
\mathbf{\hat{x}}_0(\mathbf{x}_t, \mathbf{c}, t) = D_\theta([\mathbf{z}_t \oplus \mathbf{c}], t).
\end{equation}
Here, in U-net-style network, encoder extracts features from noisy mask: $\mathbf{z}_t=\text{Enc}(\mathbf{x}_t, t)$; \(\oplus\) denotes a feature fusion operation (element-wise addition) integrating feature of noisy input $\mathbf{z}$ and PVT embeddings $\mathbf{c}$; \(D_\theta\) is the decoder.

\begin{figure}[!h]
\centering
\begin{minipage}[t]{0.48\textwidth}
\centering
\begin{algorithm}[H]
\caption{Training (PVT-conditioned)}
\begin{algorithmic}[1]
\Require Conditioning image $I$, corresponding mask $\mathbf{x}_0$
\For{each training step}
    \State $\mathbf{x}_0 \sim q(\mathbf{x}_0)$ 
    \State $t \sim \mathcal{U}\{1, \dots, T\}$ 
    \State $\epsilon \sim \mathcal{N}(0, \mathbf{I})$ 
    \State $\mathbf{x}_t \gets \sqrt{\bar{\alpha}_t} \mathbf{x}_0 + \sqrt{1 - \bar{\alpha}_t} \epsilon$ 
    \State $\mathbf{\hat{x}}_0 \gets D_\theta([\text{Enc}(\mathbf{x}_t, t) \oplus \mathcal{E}_{\text{PVT}}(I,\mathbf{x}_t, t)], t)$ \Comment{Key step}
    \State $\mathcal{L} \gets \mathcal{L}_{\dagger}(\mathbf{\hat{x}}_0, \mathbf{x}_0)$ 
    \State Update $\theta$ using $\nabla_\theta \mathcal{L}$
\EndFor
\end{algorithmic}
\label{Alg1}
\end{algorithm}
\end{minipage}
\begin{minipage}[t]{0.48\textwidth}
\centering
\begin{algorithm}[H]
\caption{Sampling (PVT-conditioned)}
\begin{algorithmic}[1]
\Require Conditioning image $I$
\State $\mathbf{x}_T \sim \mathcal{N}(0, \mathbf{I})$ 
    \State $L \gets []$
\For{$t = T, T-1, \dots, 1$}
    \State $\mathbf{\hat{x}}_0 \gets D_\theta([Enc(\mathbf{x}_t, t) \oplus \mathcal{E}_{\text{PVT}}(I,\mathbf{x}_t, t)], t)$ \Comment{Key step}
    \State $ L \gets L \mathbin{\|}\mathbf{\hat{x}}_0$
    \State $\boldsymbol{\mu}_\theta \gets \frac{\sqrt{\alpha_t} (1 - \bar{\alpha}_{t-1})}{1 - \bar{\alpha}_t} \mathbf{x}_t + \frac{\sqrt{\bar{\alpha}_{t-1}} (1 - \alpha_t)}{1 - \bar{\alpha}_t} \mathbf{\hat{x}}_0$ 
    \State $\mathbf{x}_{t-1} \gets \boldsymbol{\mu}_\theta$
    \If{$t>1$}
        \State $\mathbf{n} \sim \mathcal{N}(0, \mathbf{I})$
        \State $\mathbf{x}_{t-1} \gets \mathbf{x}_{t-1} + \sigma_t \mathbf{n}$
    \EndIf
\EndFor
\State $\mathbf{x}_0\gets \text{Consensus(L)}$
\State \Return $\mathbf{x}_0$
\end{algorithmic}
\label{Alg2}
\end{algorithm}
\end{minipage}
\end{figure}

\subsection{Profiling Protocol: Lightweight}
To assess the efficiency of our model, we benchmark on an NVIDIA A100 GPU with batch size 1 and resolution 352:
\begin{itemize}
    \item \textbf{Trainable parameters (M)}: Total number of learnable parameters reported in millions.
    \item \textbf{Reserved VRAM usage (MB)}: The maximum GPU memory reserved during training, reported by pytorch.
    \item \textbf{Typical VRAM usage (MB)}: The GPU memory utilized during training, reported by nvidia-smi command.
    \item \textbf{Training time (ms)}: The mean wall-clock time required for each forward and backward pass in training.
    \item \textbf{Inference speed (ms/image)}: The average time taken to perform a forward pass on a single image.
\end{itemize}
Here, we record reserved VRAM since Out-of-Memory errors can still occur even when the typical VRAM remains below the GPU’s capacity. This discrepancy arises because nvidia-smi has a limited reporting frequency and may fail to capture transient memory spikes.

\section{Experiments}
\label{sec:exp}
\begin{table}[!ht]
\caption{Benchmark results on three datasets. Best scores \textbf{bolded}. 
$\uparrow$ indicates higher is better, $\downarrow$ indicates lower is better.}
\label{tab:multi_dataset_benchmark}
\centering
\resizebox{0.7\linewidth}{!}{%
\begin{tabular}{c|cccc}
\toprule
\multirow{2}{*}{\textbf{Method}} & \multicolumn{2}{c}{\textbf{AbdominalCT-1K}}& \multicolumn{2}{c}{\textbf{BraTs}}\\
\cmidrule(lr){2-3} \cmidrule(lr){4-5}
& F-1 ↑ & mIoU ↑ & F-1 ↑ & mIoU ↑ \\
\midrule
\textbf{Ours}    & \textbf{93.3} & \textbf{88.3} & \textbf{84.713} & \textbf{74.79} \\
U-KAN            & 90.6          & 84.0          & 81.55    & 70.60          \\
nnUNet           & 92.4          & 87.2          & 82.657 & 72.21  \\
DSSAU-Net        & 92.6          & 87.3          & 84.711          & 74.77          \\
\bottomrule
\end{tabular}
}
\end{table}
\begin{figure}[!ht]
    \centering
    \setlength{\tabcolsep}{2pt} 
    \renewcommand{\arraystretch}{1.0}
    \begin{tabular}{c c c}
        \begin{tabular}{c}
            \rotatebox{90}{\includegraphics[width=0.28\linewidth]{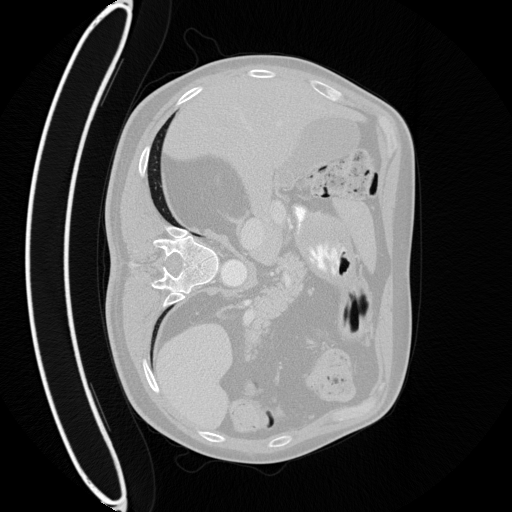}} \\
            \scriptsize (a) Image
        \end{tabular} &
        \begin{tabular}{c}
            \rotatebox{90}{\includegraphics[width=0.28\linewidth]{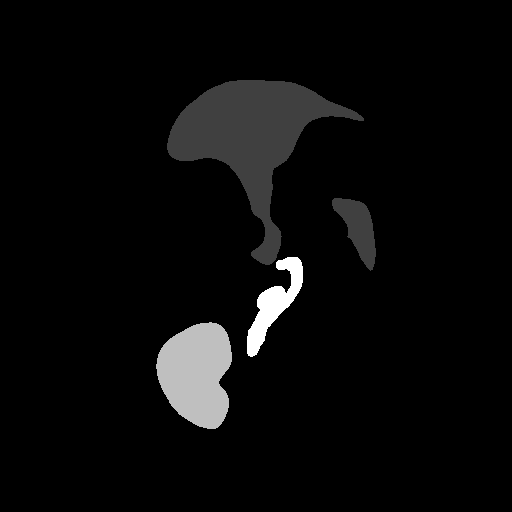}} \\
            \scriptsize (b) GT
        \end{tabular} &
        \begin{tabular}{c}
            \rotatebox{90}{\includegraphics[width=0.28\linewidth]{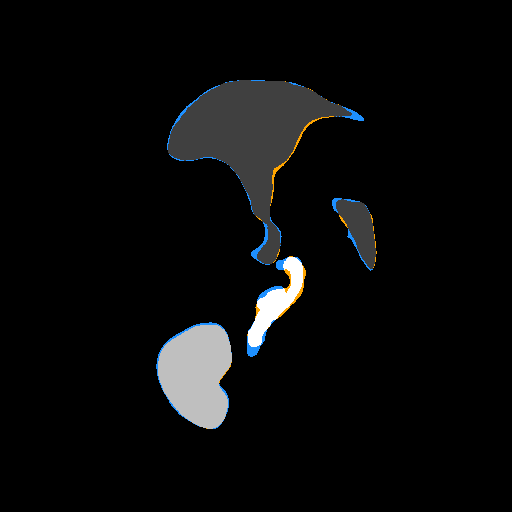}} \\
            \scriptsize (c) Ours
        \end{tabular} \\
        \begin{tabular}{c}
            \rotatebox{90}{\includegraphics[width=0.28\linewidth]{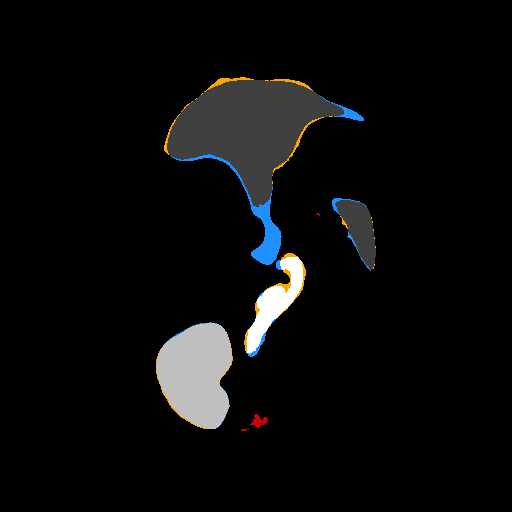}} \\
            \scriptsize (d) nnUNet
        \end{tabular} &
        \begin{tabular}{c}
            \rotatebox{90}{\includegraphics[width=0.28\linewidth]{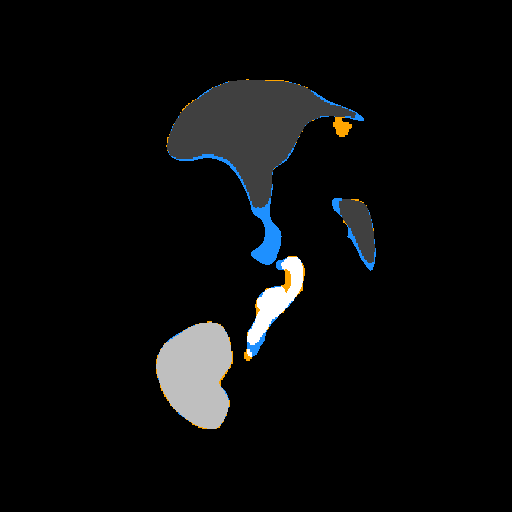}} \\
            \scriptsize (e) DSSAU-Net
        \end{tabular} &
        \begin{tabular}{c}
            \rotatebox{90}{\includegraphics[width=0.28\linewidth]{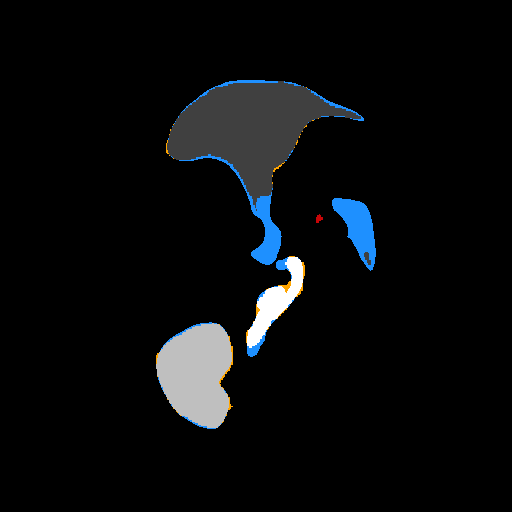}} \\
            \scriptsize (f) U-Kan
        \end{tabular} \\
    \end{tabular}
    \caption{Qualitative comparisons. MedCondDiff yields more accurate predictions with finer details and contours (Blue: false negative; Orange: false positive; Red: Hallucination).}
    \label{fig:quali-compare}
\end{figure}

\subsection{Experimental Setup}
\noindent \textbf{Datasets:} We evaluated our method on Abdominal CT 1K~\cite{Ma2021AbdomenCT1K} and MSD Brain task ~\cite{simpson2019MSD}, which includes volumetric abdominal CT and MRI and multi-modal brain MRI.

\noindent \textbf{Evaluation Metrics:} Performance is assessed using Dice Score (F1) and mean Intersection over Union (mIoU).

\noindent \textbf{Implementation Details:} Implemented in PyTorch and trained on one A100 GPU, MedCondDiff addresses class imbalance with a weighted sampler and a hybrid Dice–cross-entropy loss. The model uses a pretrained PVT-B1 backbone~\cite{pvt-pretrained} and is trained for 10 epochs with AdamW, an initial learning rate of 1e-4, and batch size 16.

\subsection{Results}
To ensure fairness, we sample four segmentation masks per test case and select the best for visualization. Fig.~\ref{fig:quali-compare} shows representative results against baselines, quantitative results shown in Table~\ref{tab:multi_dataset_benchmark}. MedCondDiff yields anatomically consistent masks with sharper boundaries and stronger structural fidelity across modalities.

\subsection{Analysis: Efficient and Lightweight}
To assess practicality in resource-constrained settings, we compare MedCondDiff with SOTA segmentation backbones in GPU memory usage and inference time. Table~\ref{tab:efficiency} reports VRAM, average training and inference speed under identical settings. MedCondDiff requires less reserved VRAM while maintaining competitive inference speed among diffusion-based models. 


\begin{table}[!ht]
\centering
\caption{Model Size and Efficiency on AbdominalCT-1K \\ (Best Two results Bolded)}
\label{tab:efficiency}
\resizebox{0.9\linewidth}{!}{
\begin{tabular}{lr|cc|cc}
\toprule
\multirow{2}{*}{\textbf{Model}} & \multirow{2}{*}{\textbf{Parameter \#.}} & 
\multicolumn{2}{c|}{\textbf{Memory (mb)}} &
\multicolumn{2}{c}{\textbf{Time (ms/image)}}\\
\cmidrule(lr){3-4} \cmidrule(lr){5-6}
& & \textbf{Reserved} & \textbf{Typical} & \textbf{Train} & \textbf{Sample} \\
\midrule
\textbf{Ours} & \textbf{24,651,429} & \textbf{1559} & \textbf{1557} & \textbf{46.8} & \textbf{14.8} \\
U-KAN         & \textbf{6,356,757} & 2316 & \textbf{1229} & \textbf{46.8} & 17.2 \\
nnUNet        & 46,330,627 & 4134 & 2439 & \textbf{31.1} & \textbf{9.08} \\
DSSAU-Net     & 29,251,781 & \textbf{1862} & 2387 & 193.7 & 65.83 \\
\midrule
MedSegDiff    & 129,412,554 & \multicolumn{2}{c|}{N/A}& \multicolumn{2}{c}{N/A}\\
SegDiff       & 164,163,973 & \multicolumn{2}{c|}{N/A}& \multicolumn{2}{c}{N/A}\\
\bottomrule
\end{tabular}
}
\end{table}
We report parameter sizes of several diffusion-based segmentation networks but omit their performance results. Models such as MedSegDiff \cite{wu2023medsegdiff} and SegDiff \cite{amit2022segdiff} have over five times more parameters than MedCondDiff and, when trained under our setup (1 A100 GPU, 10 epochs), do not produce statistically meaningful results. Other implementations, like DiffusionInst \cite{gu2022DiffusionInst} and Diff-UNet \cite{xing2023diffunet}, used eight and four A100 GPUs, in respectively, to reproduce reported outcomes in their paper. These computational demands make direct accuracy comparisons infeasible.

\subsection{Ablation Study}
We evaluate the proposed conditioning mechanism with three variants. The baseline is a vanilla DDPM without conditioning. PVT-Ac conditions via \textit{additive} fusion, and PVT-Cc uses \textit{concatenation}. Shown in Table~\ref{tab:ablation}, conditioning improves DDPM segmentation, with PVT-Ac boosting accuracy by 15\% and PVT-Cc by 5\% over the baseline. PVT-Ac outperforms PVT-Cc because concatenation amplifies the conditioning signal, causing later layers to overfit appended features and neglect earlier representations. We excluded cross-attention conditioning \cite{wu2023medsegdiff} due to its tendency to overfit on limited medical data. 
\begin{table}[!h]
\caption{Conditional strategy ablation on AbdominalCT-1K (PVT-Ac: additive, PVT-Cc: concatenated)}
\centering
\resizebox{0.6\linewidth}{!}{
\begin{tabular}{cc|cc}
\toprule
\textbf{PVT-Ac} & \textbf{PVT-Cc} & \textbf{F-1} & \textbf{mIoU} \\
\midrule
              &              & 77.0 & 63.1\\
              & \checkmark   & 82.4 & 70.5\\
\checkmark    &              & \textbf{93.3} & \textbf{88.3} \\
\bottomrule
\end{tabular}
}
\label{tab:ablation}
\end{table}

\section{Conclusion}
\label{sec:conclusion}
In this work, we present MedCondDiff, a conditional diffusion framework for multi-organ, multi-modality segmentation. By integrating a lightweight adapter with a PVT backbone, our model incorporates structured semantic priors into the denoising process, yielding anatomically consistent masks. We have shown that MedCondDiff delivers competitive accuracy with lower VRAM and faster inference, while robust enough to generalize across modalities without extensive fine-tuning.

\bibliographystyle{IEEEtran}
\bibliography{strings}

\end{document}